\documentclass[12pt,preprint]{emulateapj}

\usepackage{amsmath, amssymb, latexsym}

\newcommand{\Msun}{\textrm{M}_{\odot}}
\newcommand{\etal}{et al.}
\newcommand{\Mbh}{M_{\bullet}}

\begin{document}

\title{Event Rate for Extreme Mass Ratio Burst Signals in the Laser
  Interferometer Space Antenna Band}

\author{Louis J. Rubbo, Kelly Holley-Bockelmann, and Lee Samuel Finn}
\affil{Center for Gravitational Wave Physics, Pennsylvania State
  University, University Park, PA 16802}

\begin{abstract}
  Stellar mass compact objects in short period ($P\lesssim10^3$~s)
  orbits about a $10^{4.5}$--$10^{7.5}\,\rm{M}_{\odot}$ massive black
  hole (MBH) are thought to be a significant continuous-wave source of
  gravitational radiation for the ESA/NASA Laser Interferometer Space
  Antenna (LISA) gravitational wave detector.  These extreme
  mass-ratio inspiral sources began in long-period, nearly parabolic
  orbits that have multiple close encounters with the MBH.  The
  gravitational radiation emitted during the close encounters may be
  detectable by LISA as a gravitational wave burst if the
  characteristic passage timescale is less than $10^5$~s.  Scaling a
  static, spherical model to the size and mass of the Milky Way bulge
  we estimate an event rate of $\sim\!15~\textrm{yr}^{-1}$ for such
  burst signals, detectable by LISA with signal-to-noise ratio greater
  than five, originating in our Galaxy.  When extended to include
  Virgo Cluster galaxies our estimate increases to a gravitational
  wave burst rate of $\sim\!18~\textrm{yr}^{-1}$.  We conclude that
  these extreme mass-ratio burst sources may be a steady and
  significant source of gravitational radiation in the LISA data
  streams.
\end{abstract}


\section{Introduction} \label{sec:intro}

The inspiral of compact objects onto massive black holes (MBHs) in
galactic nuclei is an anticipated important gravitational radiation
source for the ESA/NASA Laser Interferometer Space Antenna (LISA)
\citep{hils:1995:gac, sigurdsson:1997:csm, glampedakis:2002:zwe,
glampedakis:2002:ait, ivanov:2002:frc, freitag:2003:gws,
gair:2004:ere, danzmann:2003:ltc, sumner:2004:ogw, jennrich:2005:pil,
seto:2001:pdm}.  These extreme mass-ratio inspiral (EMRI) sources will
complete $>10^4$ orbits during the LISA mission lifetime
($\sim\!5$~yrs) and it is the accumulation of the signal power,
emitted continuously with frequency $\gtrsim10^{-3}$~Hz, that makes
them visible to LISA.

However, before emerging as EMRIs these objects were on long-period,
nearly radial orbits that were ``captured'' by the MBH after the
orbital energy was reduced by a series of short, intense bursts of
gravitational radiation emitted during each close encounter with the
MBH.  If the encounter timescale is less than $\sim\!10^5$~s the
gravitational wave burst will be in the LISA band and, if strong
enough, detectable by LISA.  Here we describe a preliminary analysis
of this extreme mass ratio burst (EMRB) phenomenon, characterizing the
detectability and estimating the rate of EMRB events in LISA.  Our
estimates show that the Milky Way may be responsible for $\sim\!15$
observable EMRB events per year, and that the Virgo Cluster galaxies
may be responsible for an additional $\sim\!3$ events per year, making
EMRBs a significant source of gravitational waves for LISA.

\section{Inspirals vs. Bursts} \label{sec:process}

To illustrate the evolution from EMRB to EMRI consider a $0.6~\Msun$
white dwarf in a nearly radial Keplerian orbit about a MBH of mass
$\Mbh\sim10^{6}~\Msun$. If the initial orbit has an apocenter of
100~pc and a pericenter of 0.2~AU (corresponding to 10 Schwarzschild
radii for the MBH), then during the first pericenter pass the orbit
loses enough energy via gravitational radiation that the next
apocenter is $\sim\!60$~pc, while the change in pericenter is
negligible.  With each subsequent pericenter pass, a strong burst of
gravitational radiation reduces the subsequent apocenter and the orbit
becomes more circular. LISA is sensitive to gravitational waves in the
approximately $10^{-5}$ to $10^{-1}$~Hz band, so we classify as EMRBs
those systems with orbital periods greater than $10^5$~s and with
pericenter passage timescales of less than $10^5$~s. Conversely, EMRIs
have orbital periods less than or on order $10^3$~s and radiate
continuously in the LISA band.

EMRIs and EMRBs sample different components of a galactic nucleus.
Every EMRI source was, in its past, a \emph{possible} EMRB source:
i.e., it was on a highly eccentric orbit that, owing to gravitational
wave emission, circularized and decayed until it entered the LISA band
as a continuous source. However, if the MBH encounter timescale is too
long, the orbit may become an EMRI without any significant radiation
burst in the LISA band.

Not all EMRBs evolve to become EMRIs, either. An object on a highly
eccentric orbit may be scattered by other stars after encountering the
MBH, or may plunge directly into the MBH.  In addition, while massive
main sequence stars will be disrupted long before appearing as EMRIs,
the periapsis of an EMRB orbit can be much greater than the tidal
disruption radius for low-mass main sequence stars.  For example, an
M2V star could pass within 0.65~AU of the Milky Way's MBH without
disruption and, if on a nearly radial orbit so that its pericenter
velocity was highly relativistic, may radiate a gravitational wave
burst with a characteristic frequency of several mHz.

\section{Stellar Model} \label{sec:model}

To evaluate the EMRB event rate we begin with a model of our Galaxy
center.  The Milky Way bulge consists of a $\Mbh =
3.7\times10^6~\Msun$~MBH \citep{ghez:2005:soa} embedded in a stellar
ellipsoid \citep[e.g.][]{binney:1997:psi, stanek:1997:mgb,
lopez-corredoira:2000:iss} of mass $(1.3$--$2)\times10^{10}\,\Msun$
\citep[e.g.][]{gerhard:2002:mdo}, with axis lengths of
$\sim\!1.8:0.7:0.5$~kpc \citep{bissantz:2002:sab}, and with a cuspy
density profile that tends toward $\rho \sim r^{-1.8}$ as $r
\rightarrow 0$ \citep{matsumoto:1982:boc}.  Although the Milky Way is a
barred galaxy, we treat the bulge as spherical with a density profile
described by an $\eta$ model \citep{tremaine:1994:fms}. Since $\eta$
models are static, spherical, and isotropic their distribution
function depends only on energy, making the model an ideal first
choice to analytically estimate the number of bursts in the bulge.  We
choose $\eta=1.25$ to match the observed slope of the Milky Way inner
cusp and set the effective radius equal to 2~kpc, total stellar mass
to $2 \times 10^{10}~\Msun$, and MBH mass to $4 \times 10^6~\Msun$.

Embedding an $\eta$ model with a pre-existing central MBH leaves the
stellar density profile unchanged, but changes the cluster potential
so that $\Phi_{\bullet}(r) \equiv \Phi_{\star}(r) - \Mbh/r$, where
$\Phi_{\bullet}$ is the potential of the MBH-embedded model and
$\Phi_{\star}$ is the stellar potential of an $\eta$ model.  The new
potential naturally adjusts the distribution function, which can be
obtained from a spherical, isotropic density profile via Eddington's
formula \citep{binney:1987:gd}, which for a MBH-embedded model can be
solved analytically near the MBH:
\begin{equation}
  f(\epsilon) = \frac{\eta \Gamma(4-\eta)}{2^{7/2} \pi^{5/2}
    \Mbh^{(3-\eta)} \Gamma(5/2-\eta)} \; \epsilon^{3/2-\eta} \,,
\end{equation}
where $\epsilon$ is the absolute value of the energy per unit mass.
When normalized to the total mass of the model,
$f(\epsilon)d{\boldsymbol r} d{\boldsymbol v}$ is the mass contained
in the local volume element centered on ${\boldsymbol r}, {\boldsymbol
v}$.

\section{Burst Waveforms} \label{sec:waveforms}

Orbits that may generate observable EMRBs are characterized by large
orbital periods ($\sim\!1 - 10^{9}$~yrs) and high
eccentricity. Objects on these orbits may be highly relativistic
during periapsis passage, with velocities an appreciable fraction of
the speed of light ($v_{p} \sim 0.3c$).  The gravitational wave
emission from such a ``gravitational bremsstrahlung'' event emerges as
a burst, lasting as long as the MBH encounter and beamed in the
direction of the secondary's velocity at periapsis.  For this
preliminary exploration we treat trajectories near periapsis as
Keplerian,\footnote{Although we only considered elliptical systems
within the $\eta$ model, it is worth noting that the Keplerian
equivalent orbit near the MBH is not necessarily elliptical.} with the
same $v_{p}$, $r_{p}$, and mass enclosed within $r_{p}$ as dictated by
the $\eta$ model, and we calculate the gravitational radiation as if
it were well described by the quadrupole formula.  A more detailed
treatment would include the effect of beaming, the radiative
contributions from terms in $v/c$ higher than quadrupole, and take
into account the actual relativistic trajectory, all of which may be
important for trajectories with small $r_{p}/\Mbh$ or large $v/c$
\citep[cf.][]{gair:2005:sagr}.

From the (quadrupole) waveforms we calculate the signal-to-noise ratio
(SNR) in the LISA detector assuming the source is at a distance of
8~kpc. In general, the square of the SNR is given by,
\begin{equation} \label{eq:snr_defined}
  \rho^{2} = 4 \int_{0}^{\infty}
  \frac{|\widetilde{h}(f)|^{2}}{S_{n}(f)} \, df \,,
\end{equation}
where $\widetilde{h}(f)$ is the Fourier transform of the gravitational
wave signal projected onto the LISA detector and $S_{n}(f)$ is the
one-sided instrument noise power spectral density.  For our purposes,
the values of $S_{n}(f)$ are taken from the \textit{Online Sensitivity
Curve Generator} \citep{larson:2000:scs} with the standard LISA
settings and the inclusion of a white dwarf background ``noise''
contribution \citep{bender:1997:cnl}.

To gain insight into the magnitude and scaling of the SNR,
equation~\ref{eq:snr_defined} can be approximated by taking advantage
of the simple burst structure in the waveforms, which allows us to
quickly approximate the integral and Fourier transform,
\begin{eqnarray}
  \rho &\approx& 100 \left( \frac{R}{8~\textrm{kpc}} \right)^{-1} \left(
  \frac{M_{\star}}{\Msun} \right) \bigg( \frac{v_{p}}{0.3 c}
  \bigg)^{2} \left( \frac{f_{\star}}{1~\textrm{mHz}} \right)^{-1/2}
  \nonumber\\
  && \times \left( \frac{S_{n}(f_{\star}/1~\textrm{mHz})}{4 \times
  10^{-37}~\textrm{Hz}^{-1}} \right)^{-1/2} \;,
\end{eqnarray}
where $R$ is the distance to the EMRB, $M_{\star}$ is the mass of the
secondary, and $f_{\star}= v_{p}/r_{p}$ is the inverse of the burst
duration.  The value of $f_{\star}$ can be viewed as a characteristic
frequency for an EMRB event, but these events are extremely broadband
in nature.  Note that the noise spectral density scales as $f^{-4}$
for frequencies below 1~mHz and the pre-factor accounts for the white
dwarf binary background.

\section{Event Rates} \label{sec:results}

To arrive at an EMRB event rate estimate, we divided our $\eta$ model
into five million elements in $(r,v)$ phase space.  We chose the
element centers uniformly spaced in $\log(r)$ and $\log(v)$ with
$r\in[7.6\times10^{-7},2\times10^4]$~pc and
$v\in(0,2\times10^5)~\textrm{km s}^{-1}$.  For each phase-space
element we calculated $f(\epsilon)$, the average orbital period
$P_{orb}$, periapsis $r_{p}$, and velocity at periapsis $v_{p}$.

To arrive at the number of compact objects and low mass main sequence
stars in our model we do the following.  Assuming all of the mass
density in our model results from a single burst of star formation
10~Gyr ago, and that the number of stars per unit mass follows a
Kroupa IMF\footnote{We varied the IMF choice to include Salpeter and
Scalo IMFs; this did not change our rate significantly.}
\citep{kroupa:2001:vim}, we converted $f(\epsilon)$ into the total
number of stars per mass per phase space element.  This number can be
less than one per phase space element.  To determine the number of
main sequence stars per phase space element we calculated the maximum
stellar mass that would survive tidal disruption by the MBH
\citep{sridhar:1992:ic} and integrate the IMF up to this mass, setting
the main sequence stellar mass equal to the average mass that survives
tidal disruption.  We determined the white dwarf population by
assuming white dwarfs formed from all stars between $1-8~\Msun$ and
the final white dwarf mass is $0.6~\Msun$.  Similarly, neutron stars
were assigned masses of $1.4~\Msun$ and form from stars $8-20~\Msun$,
while black holes are $10~\Msun$ objects formed from $20-120~\Msun$
stars.  Our calculation neglects the effects of mass segregation,
multiple episodes of star formation, and forces all stars between
$1-120~\Msun$ to be a stellar remnant.

Only a small fraction of the initial phase space contain EMRBs. To
isolate the EMRB orbits we made a series of cuts.  We immediately
discarded those orbits that were either unbound or had periapsis
within four Schwarzschild radii.  We excluded orbits with $P_{orb}$
greater than the relaxation time, $t_{relax} \equiv v^2/D(\Delta
v_\parallel^2)$, where $D(\Delta v_\parallel^2)$ is the Fokker-Planck
diffusion coefficient \citep[eq.~8-68]{binney:1987:gd}.  We also
excluded orbits with $P_{orb}$ less than $10^{5}$~s, which would yield
a continuous signal in the LISA band and are more accurately
classified as EMRIs.  We also excluded orbits that radiate so strongly
that they plunge into the MBH faster than a dynamical time, where the
plunge timescale $T_{plunge}$ is
\begin{eqnarray}
  T_{plunge} &\approx& 3.2\times 10^{6}~\textrm{yrs} \left(
  \frac{\Mbh}{10^{6} \Msun} \right)^{2} \left( \frac{M_{\star}}{1
  \Msun} \right)^{-1} \nonumber\\
  & & \times \left( \frac{r_{p, i}}{10 R_{s}} \right)^{4}
  \left( \frac{1 - e_{i}}{10^{-5}} \right)^{-1/2} \;.
\end{eqnarray}
Here $r_{p,i}$ is the initial pericenter distance and $e_{i}$ is the
initial eccentricity.  Finally, we excluded all orbits with an
encounter timescale $\Delta t = r_{p}/v_{p} > 10^5$~s, as these passes
emit radiation outside the LISA band. The remaining stars constituted
our potential EMRB population.

\begin{figure}
\epsscale{0.9}
\plotone{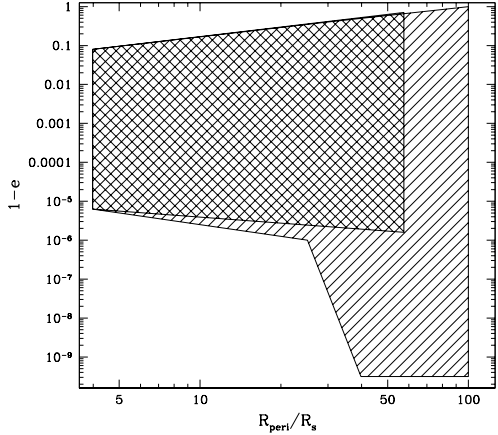}
\caption{Orbital parameters of potential EMRB sources.  Here $e$ is
    the eccentricity of the orbit within the total bulge potential,
    and $R_{p}$ is in units of Schwarzschild radii. Typical sources
    are more eccentric and have larger pericenter distances than the
    average EMRI. Overlaid is the subset of white dwarf orbits that
    are observable with LISA at a SNR~$> 5$.}
\label{fig:parmspace}
\end{figure}

Figure~\ref{fig:parmspace} shows the phase space remaining after these
cuts; $\sim\!1,800$ phase space elements remained, which does not
imply 1800 distinct objects; in fact there are typically $10^{-7}$
white dwarfs per $d{\boldsymbol r} d{\boldsymbol v}$ in this region.
Overlaid in the same figure is the subset of the white dwarf phase
space with SNR~$>5$.

\begin{figure}
\plottwo{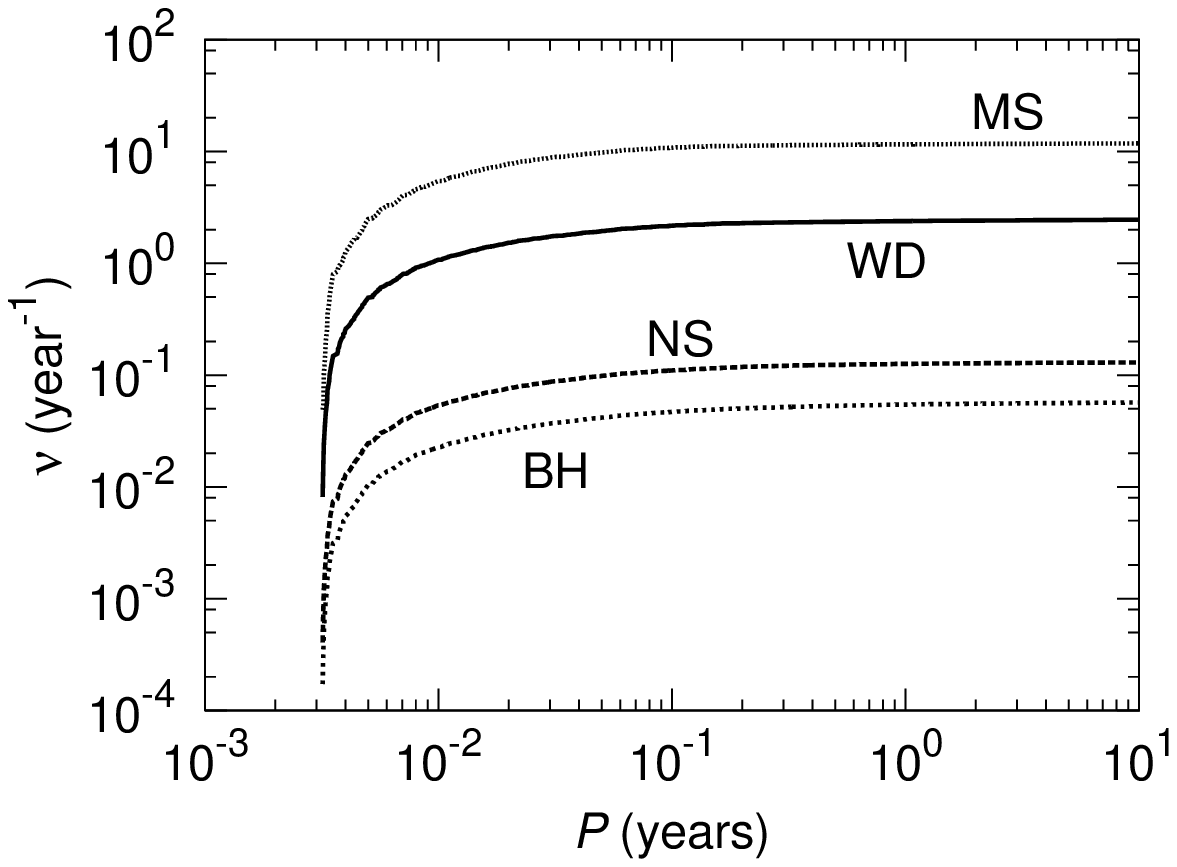}{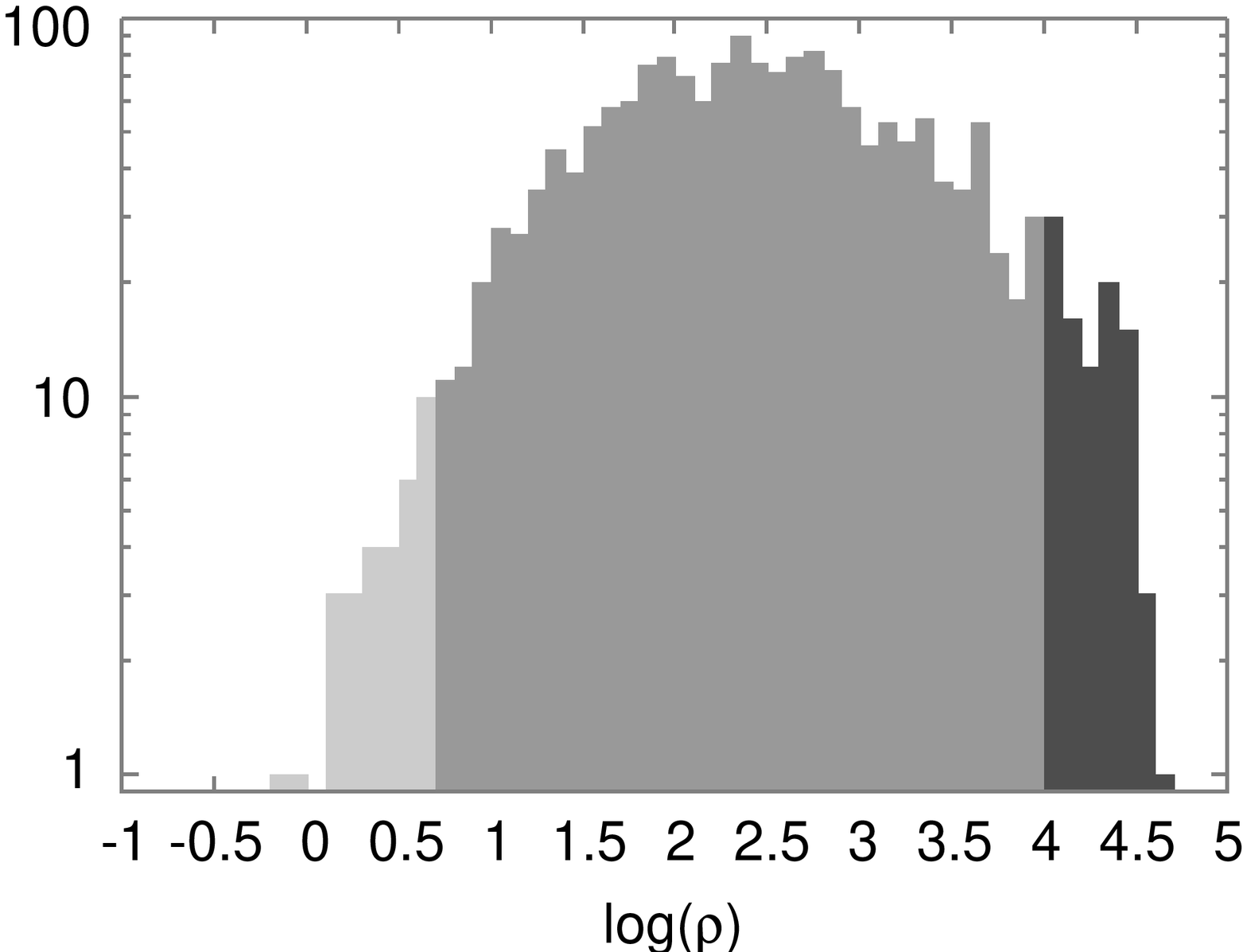}
\caption{Left: Accumulative event rate as a function of the orbital
    period.  MS = main sequence; WD = white dwarf; NS = neutron star;
    BH = black hole.  Right: Number of black hole sources versus
    signal to noise ratio.  All sources above the light grey section
    have SNR $>5$. Note the small fraction of EMRBs with SNR$>10^{4}$
    (dark grey), which would be visible to Virgo distances. }
\label{fig:TauVsP}
\end{figure}

The total event rate $\nu$ is the sum over all orbits with LISA SNR
greater than 5:
\begin{equation}
  \nu = \sum_{\rho > 5} \left(
  \frac{N_{\textrm{LMMS,survive}}}{P_{\textrm{orb}}} + 
  \frac{N_{\textrm{WD}}}{P_{orb}}+
  \frac{N_{\textrm{NS}}}{P_{orb}} + 
  \frac{N_{\textrm{BH}}}{P_{orb}} \right) \,.
\end{equation}
For our Milky Way bulge model, we find $\nu_{\textrm{WD}} =
3~\textrm{yr}^{-1}$, $\nu_{\textrm{NS}} = 0.1~\textrm{yr}^{-1}$,
$\nu_{\textrm{BH}} = 0.06~\textrm{yr}^{-1}$, and
$\nu_{\textrm{LMMS,survive}} = 12~\textrm{yr}^{-1}$.  These event
rates imply that EMRBs may be an important, heretofore unrecognized
source of gravitational waves in the LISA band.  The left panel of
figure~\ref{fig:TauVsP} shows the accumulative event rate as a
function of orbital period.  Most EMRBs originate from orbits that
would produce multiple bursts of radiation per year in the LISA band.
The right panel of figure~\ref{fig:TauVsP} shows the number of black
hole EMRBs as a function of SNR; note that some have SNRs high enough
to be seen out to Virgo distances.  If we repeat our analysis with the
EMRB distances set to 16~Mpc and multiply by the number of galaxies in
the cluster with suitable mass MBHs, we find a Virgo event rate of
$\sim\!3~\textrm{yr}^{-1}$, all due to encounters of low mass black
holes with central MBHs.

\section{Discussion} \label{sec:discussion}

EMRBs may comprise a significant new low frequency gravitational wave
source, rivaling the total number of EMRI events visible in any year
of LISA observation. For example, \citet{freitag:2001:mcc} estimates
that $\mathcal{O}(0.1)$ white dwarfs per year are in the EMRI phase in
the Milky Way. We find $\mathcal{O}(1)$ EMRBs per year from close
white dwarf encounters. The relaxation requirements on EMRBs are much
less strict: EMRIs require thousands of undisturbed orbits to
accumulate a SNR large enough to be detected ($T_{plunge} > (1-e)
T_{relax}$), while EMRBs merely need to pass by the MBH before
scattering via 2-body relaxation.

Our estimated Milky Way EMRB event rate of $\sim\!15~\textrm{yr}^{-1}$
may appear high in light of other processes, such as the white dwarf
capture rate, which estimates place $\mathcal{O}(10^{-7})$ per year
\citep{freitag:2001:mcc}; however, the greatest contribution to the
event rate we calculate arises from sources that burst multiple times
a year, yet individually have a small phase space density. In other
words, our rate does not imply that $15~\Msun$ of material interact with
the black hole per year; rather, given these EMRB orbits we calculate
that Milky Way black hole grows by less than $10^{4}~\Msun$ over a
Hubble time via EMRB decay, with many EMRBs bursting hundreds of times
before falling into the MBH. Our simple stellar model passes other
sanity-checks as well: given the capture definition of
\citet{freitag:2001:mcc}, we expect a white dwarf capture rate of
$\mathcal{O}(10^{-6})$ per year and find, in our calculations, a tidal
disruption rate of $\mathcal{O}(10^{-5})$ per year, which is
consistent with observational estimates \citep{donley:2002:lxray}.

We have introduced EMRBs as the evolutionary precursor to EMRIs; many
EMRBs will inspiral and eventually evolve into EMRIs as gravitational
radiation carries away energy and angular momentum. This implies that
the lifespan of the typical EMRB phase may only be $\mathcal{O}(10^7)
- \mathcal{O}(10^9)$ years after galaxy formation or a star formation
episode.  We have assumed, however, that EMRB phase space is always
occupied. For there to be a significant population of EMRBs in the
present-day galaxy sample, there must be a refilling mechanism.  There
are several mechanisms that may refill the burst reservoir. One such
mechanism, the dynamical migration of a stellar cluster, may have
recently left its mark in the inner 0.04 pc of our galaxy in the form
of the young S stars \citep{kim:2004:dfg}.  Other mechanisms, such as
triaxiality \citep{merritt:2004:apj, khb:2006:astroph} or resonant
relaxation \citep{hopman:2006:astroph}, can act to refill the loss
cone on timescales much shorter than the two-body relaxation time.

Since we were primarily concerned with determining whether EMRBs were
an overlooked class of LISA sources, we considered the simplest
possible model for the Milky Way bulge, the radiation properties, and
the ability of gravitational wave analysis techniques to detect the
signal.  Under these approximations the predicted event rate implies
that bright EMRBs are so numerous that the local universe will
generate a background of gravitational wave bursts with a mean rate of
once per three weeks. Changing the assumptions, however, can alter
this rate by several orders of magnitude; modeling the bulge as a bar,
for example, can increase the capture rate by two orders of magnitude
\citep{khb:2006:astroph}, while mass-segregation can {\em decrease}
the EMRB rate from low mass main sequence stars by two orders of
magnitude \citep{freitag:2003:gws}.

The relatively large event rate calculated here suggests that EMRBs
may in fact be a significant source for the LISA detector and
indicates that further, more accurate modeling needs to be undertaken.
To better characterize EMRBs and study how they can constrain models
of galactic nuclei, future work needs to include a more accurate model
for the gravitational wave emission, a better treatment of the orbit
in the relativistic regime, more realistic time-dependent galaxy and
star formation models, and a proper treatment of stellar dynamics,
including mass segregation.  In addition, the potential presence of
these burst gravitational wave sources introduces a new set of data
analysis challenges, associated with the detection and
characterization of the EMRB signal from LISA data.  As daunting as
the prospect of these challenges are their potential payoff: the
signal from EMRBs probe the central region of our Galaxy, and galaxies
in the Virgo Cluster, at scales less than $10^{-5}$ pc, far greater
resolution than we can ever hope to achieve electromagnetically.

\acknowledgments

The authors are grateful to T. Bogdanovic, M. Freitag, P.  Laguna,
S. Larson, C. Miller, S. Sigurdsson, and D. Shoemaker for numerous
helpful discussions on all aspects of this project.  This work was
supported by NASA NNG04GU99G, NASA NNG05GF71G, NSF PHY 00-99559 and
the Center for Gravitational Wave Physics, which is funded by the
National Science Foundation under cooperative agreement PHY 01-14375.

%\clearpage

%\begin{figure}
%\plotone{f1.eps}
%\caption{Quadrupole EMRB waveforms.  The large $h_{+,\times}(t)$
%    values are attributed to the relativistic velocities associated
%    with the object's pericenter pass.}
%\label{fig:waveforms}
%\end{figure}

%\clearpage

%\clearpage

\end{document}